# Converse Magnetoelectric Experiments on a Room Temperature Spirally Ordered Hexaferrite

**Khabat Ebnabbasi and Carmine Vittoria**
Department of Electrical and Computer Engineering, Northeastern University, Boston MA 02115

**Allan Widom**
Physics Department, Northeastern University, Boston MA 02115

Experiments have been performed to measure magnetoelectric properties of room temperature spirally ordered $Sr_3Co_2Fe_{24}O_{41}$ hexaferrite slabs. The measured properties include the magnetic permeability, the magnetization and the strain all as a function of the electric field E and the magnetic intensity H. The material hexaferrite $Sr_3Co_2Fe_{24}O_{41}$ exhibits broken symmetries for both time reversal and parity. The product of the two symmetries remains unbroken. This is the central feature of these magnetoelectric materials. A simple physical model is proposed to explain the magnetoelectric effect in these materials.

## INTRODUCTION

There has been considerable recent interest in the nature of magnetoelectric materials[1]. Of present interest are spirally ordered hexaferrites[2-5] which have strong magnetoelectric effects at room temperature. Neutron scattering experiments performed in [5,6] revealed a spiral spin configuration responsible for the magnetoelectric effect at room temperature in $Sr_3Co_2Fe_{24}O_{41}$ hexaferrite. $Sr_3Co_2Fe_{24}O_{41}$ is identified as a Z-type hexaferriteconsisting of S,R and T "spinel" blocks (see [4,5] for details of the crystal structure). It was further revealed that in the T block the Fe-O-Fe bond angles were slightly deformed to affect the super-exchange interaction between the Fe ions and induce the spiral spin configuration in $Sr_3Co_2Fe_{24}O_{41}$. The magneto-electric coupling parameter, α, was measured to be ~ $0.5\times10^{-2}$ (CGS) at room temperature which is more than 50 times larger than the α measured in $Cr_2O_3$. This is a significant result in terms of the practical implications at room temperature. Most magnetoelectric materials operate at low temperatures, such as $TbMnO_3$, CuO, $Cr_2O_3$, etc.. The magnetoelectric effect in $Sr_3Co_2Fe_{24}O_{41}$ was confirmed previously [5] by conventional measuring techniques whereby a magnetic field, H, was applied to induce electric polarization, P, and changes in the dielectric constant , $\varepsilon_r$, at low frequencies (~ 100KHz). We have adopted an unconventional technique of measuring the magnetoelectric effect in these materials by applying an electric field, E, instead and induce changes in magnetization, permeability and strain. We refer to these measurements as the "converse" magnetoelectric measurements. Although previous authors [1-5] have established a strong correlation between the spiral configuration and the magnetoelectric effect, we provide a physical picture or model for the effect. Our measurments reveal that $Sr_3Co_2Fe_{24}O_{41}$ is electrostrictive. As such, the application of E strains the material and , therefore, changing the physical structure of the spiral spin configuration. It is this physical motion of the spiral with respect to E that induces a change in magnetization. We refer to this model as the "Slinky" helix model. Our model should be contrasted with the model for the magnetoelectric effect in ferromganetic metal films whereby the band energies of the up and down spin are modified by the electric fields at the interface between ferromagnetic and ferroelectric films. The change in the splitting of the bands leads to changes of magnetization at the surface. [7].

The thermodynamic enthalpy per unit volume $\omega(s,\mathbf{E},\mathbf{H},\sigma)$ determines all of the spirally ordered hexaferrite thermodynamic equations of state[8] via

$$d\omega = Tds - \mathbf{P} \cdot d\mathbf{E} - \mathbf{M} \cdot d\mathbf{H} - e : d\sigma \quad (1)$$

Here, T, P, M, and e represent, respectively, the temperature, polarization, magnetization and strain, while s, E, H, and σ represent, respectively, the entropy per unit volume, electric field, magnetic intensity and stress. Other thermodynamic quantities of interest include the adiabatic dielectric constant tensor

$$\varepsilon = 1 + 4\pi \left(\frac{\partial P}{\partial E}\right)_{s,H,\sigma} = 1 + 4\pi \chi_P, \quad (2)$$

the adiabatic permeability tensor

$$\mu = 1 + 4\pi \left(\frac{\partial M}{\partial H}\right)_{s,E,\sigma} = 1 + 4\pi \chi_M, \quad (3)$$

and the adiabatic magnetoelectric tensor

$$\alpha = \left(\frac{\partial M}{\partial E}\right)_{s,H,\sigma} = \left(\frac{\partial P}{\partial H}\right)_{s,E,\sigma} \quad (4)$$

Conventional experiments probing magnetoelectric effects, measure elements of the magnetoelectric tensor $\alpha_{ij} = (\partial M_i / \partial E_j)_{s,H,\sigma}$. In the converse experiments reported in this work, the magnetoelectric effect is probed by measuring elements of the magnetic permeability tensor **μ** and the strain tensor **e**, while noting the manner in which these tensors depend on E and H. Direct measurements of the magnetization M were also employed.

**EXPERIMENTAL RESULTS**

  **Experimental Material Growth**
We have adopted similar procedure in preparing single phase of $Sr_3Co_2Fe_{24}O_{41}$ except for the following preparation steps. In order to prevent formation of other impurity phases, including W-, M- and/or Y- types phases, it was found most favorable to quench the sample immediately to room temperature after annealing. The resulting X-ray diffraction pattern is shown in FIG.1. Also, for the magnetoelectric measurements it is important to minimize conductance current flow or heating effects through the sample in the presence of high electric fields. As such, the resistivity was increased by annealing the samples at 600 °C in an oxygen atmosphere for another six hours.
The resistivity estimated from the experimental linear V- I characteristic was $\rho = 1.43 \times 10^9$ Ω-cm for samples with 1 millimeter thickness. The preparation in oxygen leads to $Fe^{++}$ concentration reduction which then lowered the hopping of electrons between $Fe^{++}$ and $Fe^{+++}$ ions[5, 9].
Since we are proposing here a set of converse experiments such as the measurment of permeability with frequency, it called for unconventional measuring techniques. Typically, coaxial lines are used to measure permeability and dielectric constants as a function of frequency, but never FIG. 2: Real and imaginary parts of the polycrystalline Sr Z-type permeability versus frequency. in the presence of an electric field or a DC voltage as high as 1-2000 VDC. The risks to instrumentation are too high. We have developed a coaxial line technique whereby the inner and outer conductors are coupled at RF fields, but not at DC voltages. This technique will be described elsewhere. The microwave experiments were performed under the following conditions: For a given direction of remanence magnetization, Mr the electric field was applied parallel, opposite and perpendicular to $M_r$. Prior to the experiments the remanence direction was poled with a permanent magnet. In FIG.2 we exhibit the complex relative magnetic permeability $\mu(\omega+i0^+)$ for low microwave frequencies on the scale of the ferromagnetic resonant frequency. In the limit of f→0, we expect the permeability, µ(0), to be in the order of [10]

$$\mu(0)=1+\left(\frac{4\pi M_r}{H_\varphi}\right) \qquad (5)$$

where $4\pi M_r$ is the remanence magnetization and $H\varphi$ is the six fold magnetic anisotropy field. We measured $4\pi M_r$=105 G and ,therefore, $H\varphi \approx 40$ Oe, see Fig.2.

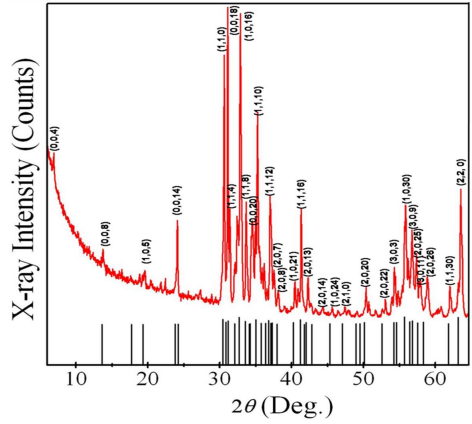

FIG. 1: X-ray di_raction pattern of the polycrystalline $Sr_3Fe_{24}Co_2O_{41}$ at room temperature. The black lines represent the reference peak positions for the Ba Z-type hexaferrite

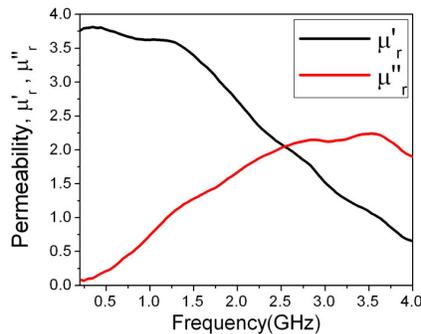

FIG. 2: Real and imaginary parts of the polycrystalline Sr Z-type permeability versus frequency.

## Experimental Magneto-Electric Measurements

Shown in FIG.3 is the change in permeability when an electric field is applied parallel or anti-parallel to the magnetization. Under a change in parity E → -E and $M_r$ → $M_r$. Under time reversal, E → E and $M_r$ → -$M_r$. The data indicates both a broken parity and a broken time reversal symmetry with the product of the symmetries unbroken. This represents the fundamental symmetry expected of magnetoelectric effects. The measurements in FIG.3 correlate very well with the vibrating sample measurements (VSM) whereby $M_r$ scales as E, changing polarity with the direction of E. Thus, in the limit as f → 0, $\Delta\mu_r$ may be explained simply in terms of changes in $M_r$ with E. However, at high frequencies dynamical magnetoelectric interaction between the RF magnetization and electric fields must be considered, since, for both directions of E, $\Delta\mu_r$ decrease monotonically with frequency. For contrast we exhibit in FIG.4, the change in permeability for electric fields perpendicular to the magnetization. The permeability shift for this case is a function only of $|E_\perp|^2$. For either case E anti-parallel or perpendicular to M the change in permeability, $\Delta\mu_r$, decreased precipitously with frequency. Simple arguments would say that $\Delta\mu_r$ to be approximately constant up to FMR frequency in zero field.

$$f = 2.8\sqrt{H_\varphi(4\pi M_r + H_\theta)} \qquad (6)$$

where $H_\varphi$=40 Oe, $4\pi M_r$=105 G, H≈20 KOe and as a result f ≈3 GHz. However, this is not the case. Clearly, there are dynamic magnetoelectric effects that need to be accounted for in order to explain the monotonic decrease of $\Delta\mu_r$ with frequency. For proper discussion of the monotonic decrease of $\Delta\mu_r$ with frequency one must include magnetic and electric relaxation effects in the Landau and Khalatnikov [11] equation pertaining to dynamic interactions in magnetoelectric materials. We will address these effects in a future publication. Finally, in FIG.5 the strain induced by an electric field is exhibited as a function of the electric field. The strain is quadratic in the electric field strength which indicates that $Sr_3Co_2Fe_{24}O_{41}$ is neither ferroelectric or piezoelectric material. Hence, the material exhibits electrostriction.

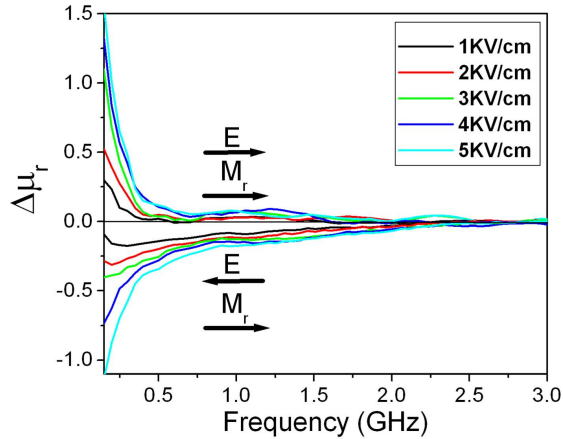

FIG. 3: The magnetic permeability change versus electric field is exhibited over a microwave frequency range when M is parallel and anti-parallel to E.

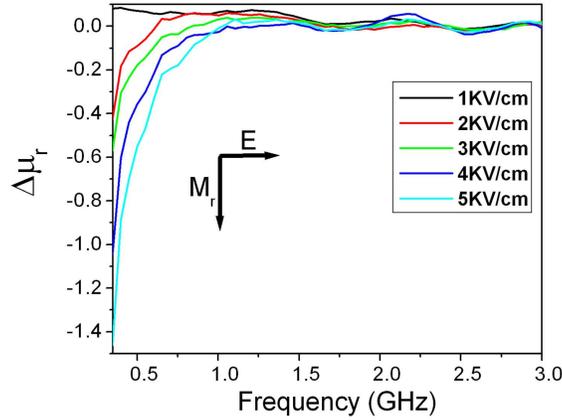
FIG. 4: The magnetic permeability change versus electric field is exhibited over a microwave frequency range when M is perpendicular to E.

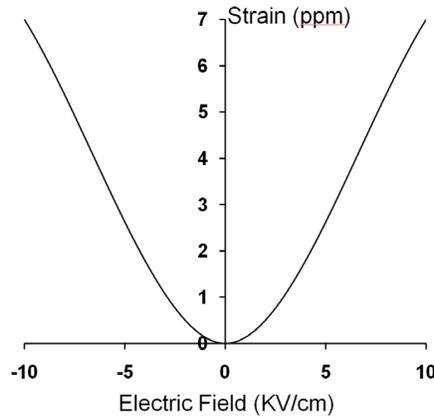
FIG. 5: The electrostriction strain of poly-crystalline Sr Z-type depends on the electric field.

## CONCLUSIONS AND DISCUSSIONS

In general terms, the magnetoelectric effect implies the following: the application of a magnetic intensity H induces a change in electric polarization P and the application of an electric field E induces a change in magnetization M. In the majority of the previous works on magneto-electric effect, the change in material parameters including the electric polarization have been measured versus magnetic field intensity only. The material hexaferrite $Sr_3Co_2Fe_{24}O_{41}$ exhibits broken symmetries for both time reversal and parity. The product of the two symmetries remains unbroken. This is the central feature of these magnetoelectric materials. Measurements have been made in order to verify this feature but in a novel manner. The measurements involve the magnetic permeability, magnetization, and strain all as a function of the electric field E and the magnetic field H. The field dependence on strain indicates that the material is electrostrictive. In our so called converse experiments we have reversed the roles of E and H whereby magnetoelectric effects were measured with the application of E rather than H. As such from practical considerations it simplifies the design of devices and applications. For example there would be less of a need for permanent magnets in microwave device applications.